# WKGM: Weight-K-space Generative Model for Parallel Imaging Reconstruction

Zongjiang Tu, Die Liu, Xiaoqing Wang, Chen Jiang, Pengwen Zhu, Minghui Zhang, Shanshan Wang, *Senior Member, IEEE*, Dong Liang, *Senior Member, IEEE*, Qiegen Liu, *Senior Member, IEEE*

*Abstract*—Deep learning based parallel imaging (PI) has made great progresses in recent years to accelerate magnetic resonance imaging (MRI). Nevertheless, it still has some limitations, such as the robustness and flexibility of existing methods have great deficiency. In this work, we propose a method to explore the k-space domain learning via robust generative modeling for flexible calibration-less PI reconstruction, coined weight-k-space generative model (WKGM). Specifically, WKGM is a generalized k-space domain model, where the k-space weighting technology and high-dimensional space augmentation design are efficiently incorporated for score-based generative model training, resulting in good and robust reconstructions. In addition, WKGM is flexible and thus can be synergistically combined with various traditional k-space PI models, which can make full use of the correlation between multi-coil data and realize calibration-less PI. Even though our model was trained on only 500 images, experimental results with varying sampling patterns and acceleration factors demonstrate that WKGM can attain state-of-the-art reconstruction results with the well-learned k-space generative prior.

## I. INTRODUCTION

Image reconstruction from undersampled measurements is an essential technique to accelerate MR imaging. However, due to the violation of the Nyquist sampling theorem, undersampling in k-space introduces aliasing artefacts. Parallel imaging (PI) [1]-[3] techniques have the ability of reconstructing uncorrupted images from undersampled measurements, with the aid of multiple receiving coils.

PI reconstruction methods can be classified broadly into two main types depending on the way in which they utilize the coil sensitivity information, namely the image domain methods and the k-space domain methods. Techniques such as sensitivity encoding (SENSE) [2], explicitly require sensitivity information for reconstruction. Theoretically, SENSE leverages the acquired sensitivity information of multiple receiving coils to remove aliasing artifacts. In practice, it is often very difficult to accurately and robustly obtain the coil sensitivity information. Another type, such as generalized autocalibrating partially parallel acquisitions (GRAPPA) [3], implicitly uses sensitivity information for reconstruction. Concretely, GRAPPA is an autocalibrating method and additional k-space data, the so-called auto calibration signal (ACS) lines, are required to estimate the GRAPPA kernels. Some other autocalibrating methods are, namely, iterative self-consistent parallel imaging reconstruction (SPIRiT) [4], auto-calibrated k-space sparse matrix (kSPA) [5], and efficient eigenvector based SPIRiT (ESPIRiT) [6]. However, once the number of acquired ACS is limited, the accuracy of estimated kernels decreases, resulting in reconstruction errors. With regard to this issue, calibrationless methods that do not require specific ACS have been proposed including calibration-less multi-coil MRI (CaLM MRI) [7], simultaneous autocalibrating and k-space estimation (SAKE) [8]; parallel-imaging low-rank matrix modeling of local k-space neighborhoods (P-LORAKS) [9] and annihilating filter-based low-rank Hankel matrix approach (ALOHA) [10]. Although promising, these reconstructions usually need much more computation for the iterative reconstruction, necessitating new methods that obviate these limitations.

Deep learning methods perform fast online reconstruction with the aid of offline training. Deep learning methods can directly explore prior knowledge either in the image domain or the k-space domain. For image domain-based approaches, Hammernik *et al.* [11] designed a Variational Network (VN)-based method to improve multi-channel MRI reconstruction. Schlemper *et al.* [12] proposed a framework for MRI reconstruction using a deep cascade of convolutional neural network to accelerate the data acquisition. Aggarwal *et al.* [13] developed a model-based deep learning architecture to achieve PI, called MoDL. Duan *et al.* [14] introduced a variable splitting network (VS-Net) to efficiently reconstruct the undersampled MR images. Liu *et al.* [15] presented a sparse operator-enhanced iterative network (IFR-Net) for fast CS-MRI reconstruction. Regarding the k-space domain-based methods, inspired by a link between the ALOHA and deep learning, Han *et al.* [16] showed a fully data-driven deep learning algorithm for k-space interpolation. All of these are the supervised learning techniques that require a great number of paired data for network training.

In recent years generative models have seen remarkable advances. Deep generative models have the advantages of alleviating the deficiency of learning flexibility. In 2014, Goodfellow *et al.* [17] first designed the generative adversarial networks (GANs). Then, various improved GAN models were proposed, which achieved superior performance in MRI. However, the training of GANs can be unstable due to the adversarial training procedure. Score-based generative models (SGMs) [18]-[20] and denoising diffusion probabilistic models (DDPMs) [21]-[22] have gained remarkably success as an emerging class of generative mod-

This work was supported in part by National Natural Science Foundation of China under 61871206, 62161026. (Corresponding authors: D. Liang and Q. Liu)

Z. Tu, D. Liu, C. Jiang, Q. Liu and M. Zhang are with School of Information Engineering, Nanchang University, Nanchang 330031, China. ({tuzongjiang, liudie, jiangchen }@email.ncu.edu.cn, {zhangminghui, liuqiegen}@ncu.edu.cn)

X. Wang is with the institute of Biomedical Imaging, Graz University of Technology, 8010, Graz, Austria. (xiaoqingwang2010@gmail.com)

P. Zhu is with College of Engineering, Pennsylvania State University, State College, PA16802, USA. (ppz5025@psu.edu)

S. Wang and D. Liang are with Paul C. Lauterbur Research Center for Biomedical Imaging, SIAT, Chinese Academy of Sciences, Shenzhen 518055, China. (sophiasswang@hotmail.com, dong.liang@siat.ac.cn)

els that have proven promising sample quality without requiring adversarial training. For instance, in the seminal work [23], Song *et al.* proposed a score-based diffusion model by looking at the stochastic differential equation (SDE) associated with the inference process. Soon after, they applied the above technique to medical image reconstruction [24]. Similarly, Quan *et al.* [25] presented HGGDP for MRI reconstruction by taking advantage of the denoising score matching. By leveraging the learned score function as a prior, Chung *et al.* [26] introduced a score-POCS model for solving inverse problems in imaging. Moreover, Jalal *et al.* [27] trained a generative prior for MRI reconstruction. Different from the above-mentioned MRI reconstruction algorithms which are mainly based on the image domain, Xie *et al.* [28] proposed a k-space domain measurement-conditioned denoising diffusion probabilistic model (MC-DDPM) for MRI reconstruction. However, it is conditioned on specific sampling strategy. In this work, we propose a weighted k-space generative model (WKGM) that is not subject to any sampling conditions. The proposed model enormously explores the k-space domain via robust generative modeling for flexible MRI reconstruction. In addition, WKGM is trained with only 500 images, while [26]-[28] is trained with tens of thousands of images.

Specifically, we train an unsupervised score-based generative model on weighted and high-dimensional k-space data, where weighting technology and high-dimensional space augmentation design are applied to the initial k-space data for better capturing the prior distribution. The k-space weighting technology can effectively reduce the dynamic range of the k-space data and the image support by suppressing low frequency and lifting high frequency, making the network more suitable for training. In the reconstruction stage, we employ Predictor-Corrector (PC) sample as a numerical SDE solver to generate samples and embed data consistency operation. In particular, the reconstruction architecture can integrate any traditional iterative algorithms (e.g., SAKE) into the iterative reconstruction process, which can make full use of the correlation between multi-coil data and realize calibration-less PI. We highlight our main contributions as follows:

- **Robust K-space Generative Model:** Weighted and high-dimensional k-space data is used to train score-based generative model. This strategy makes the model training to be more tractable and robust.
- **Flexible Reconstruction Scheme:** The generative model is utilized in the k-space domain in an iterative manner, enabling flexible incorporation of traditional k-space domain reconstruction methods to further improve the performance.

The remainder of this paper is organized as follows: Section II briefly reviews some related PI methods in k-space domain and the research of generative models especially SGMs. In Section III, we first introduce the skeleton of WKGM and then we elaborate on the motivation for using the weighted technique and high-dimensional strategy on k-space data. Additionally, we describe the prior learning and iterative reconstruction process of WKGM in detail. Finally, we introduce the combination of WKGM and traditional algorithm. Section IV presents the MRI reconstruction performance of the proposed method. Section V concludes our findings in this work.

## II. RELATED WORK

### A. Priors in k-Space PI

The observation acquisition forward model for the multi-coil parallel MRI reconstruction problem in k-space domain can be described as follows:

$$y_c = Mk_c + n_c, \ c = 1, 2, \cdots, C \quad (1)$$

where $M$ is a diagonal matrix whose diagonal elements are either 1 or 0 depending on the sampling pattern. $n_c$ is the measurement noise. $k_c$ represents the $c$-th coil k-space data and $y_c$ denotes the corresponding acquired k-space measurement data. $C$ denotes the number of coils. The PI recovery in the k-space domain can be formulated as an optimization problem:

$$\min_k \|\mathbf{M}k - y\|_2^2 + \lambda R(k) \quad (2)$$

where $\mathbf{M} = \text{diag}[M, M, \cdots, M]$, $k = [k_1, k_2, \cdots, k_C]^T$, $y = [y_1, y_2, \cdots, y_C]^T$. $\|\mathbf{M}k - y\|_2^2$ is the data fidelity term, which enforces data consistency with acquired measurements $y$. $\|\bullet\|_2^2$ represents the $l_2$-norm. $R(k)$ is the regularization term of $k$, and $\lambda$ is the factor that balances the data-fidelity term and the regularization term.

Until now, there exist some algorithms to tackle Eq. (2). For instance, Haldar developed a framework for constrained image reconstruction that used low-rank matrix modeling of local k-space neighborhoods (P-LORAKS) [9]. Based on previous works, Akċakaya *et al.* [30] further put forward robust artificial-neural-networks for k-space interpolation (RAKI) that is trained on autocalibration signal data. Meanwhile, Kim *et al.* [31] proposed a similar approach, but used a recurrent neural network model named LORAKI. Also, Arefeen *et al.* [32] designed a scan-specific model (SPARK) that estimated and corrected k-space errors made when reconstructing accelerated MR data. Extensive numerical experiments show that their proposed methods consistently provide state-of-the-art reconstruction. We attribute this success to two key advantages of k-space domain learning: On the one hand, it can avoid the distortion or disappearance of detailed structures caused by the zero-filled reconstruction. On the other hand, it avoids the conversion between image domain and k-space domain in each iteration.

### B. Generative Models for MRI Reconstruction

Deep generative models have shown the ability to produce high fidelity results in MRI reconstruction with different strategies. For example, Tezcan *et al.* [33] utilized the variational autoencoders (VAEs) as a tool for describing the data density prior. In the follow-up development, many researchers continued to carry out further research on the basis. Liu *et al.* [34] leveraged a denoising autoencoding (DAE) network as an explicit prior to address the highly undersampling MRI reconstruction problem. In [35], Bora *et al.* presented an algorithm that used generative adversarial network (GAN) for compressed sensing. Autoregressive generative models like PixelCNN have also been verified in MRI reconstruction [36]. In addition, a new generative flow (Glow) [37] formulated in the latent space of invertible neural network-based generative models was proposed for reconstructing MR images. More recently, by using energy function as the generative model, Guan *et al.* [38] proposed a new method of MRI reconstruction by utilizing a deep energy-based model. Another promising direction to address unsupervised MRI reconstruction is using score-based diffusion models. Song *et al.* [23] first presented a framework for score-based generative modeling based on stochastic differential equation (SDE). Later, Chung *et al.* [26] used the above idea to solve inverse problems in MRI, and have proven its effectiveness in accelerated MRI reconstruction.

*C. Score-based Generative Model with SDE*

Score-based generative models have gained a lot of successes in generating realistic and diverse data recently. Score-based approaches define a forward diffusion process for transforming data to noise and generating data from noise by reversing the forward process. Recently, Song *et al.* [23] presented a SDE that transforms a complex data distribution to a known prior distribution by slowly injecting noise and a corresponding reverse-time SDE that transforms the prior distribution back into the data distribution by slowly removing the noise.

More specifically, we consider a diffusion process $\{x(t)\}_{t=0}^T$ with $x(t) \in \mathbb{R}^n$, where $t \in [0,T]$ is a continuous time variable and $n$ denotes the image dimension. By choosing $x(0) \sim p_0$ and $x(T) \sim p_T$, $p_0$ to be the data distribution and $p_T$ to be the prior distribution, the diffusion process can be modeled as the solution of the following SDE:

$$dx = f(x,t)dt + g(t)dw \quad (3)$$

where $f \in \mathbb{R}^n$ and $g \in \mathbb{R}$ is the drift coefficient and the diffusion coefficient of $x(t)$, respectively. $w \in \mathbb{R}^n$ induces Brownian motion.

According to the reversibility of SDE, the reverse process of Eq. (3) can be expressed as another stochastic process:

$$dx = [f(x,t) - g(t)^2 \nabla_x \log p_t(x)]dt + g(t)d\overline{w} \quad (4)$$

where $dt$ is the infinitesimal negative time step, and $\overline{w}$ is a standard Wiener process for time flowing in reverse. The sore term $\nabla_x \log p_t(x)$ can be approximated by a learned time-dependent score model $s_\theta(x_t,t)$. The SDE is then solved by means of some solver procedure, providing the basis for score-based generative modeling with SDEs.

## III. METHOD

*A. Skeleton of WKGM*

To obtain richer information and high-quality results, the prior distribution is estimated using the score-based generative models. Specifically, instead of perturbing data with a finite number of noise distributions, a continuous distribution over time is considered with a diffusion process. The samples from the prior distribution can be obtained from the reverse SDE, which can be discretized as follows:

$$\begin{aligned} k_i &= k_{i+1} + (\sigma_{i+1}^2 - \sigma_i^2)\nabla_k \log p_t(k_{i+1}) + \sqrt{\sigma_{i+1}^2 - \sigma_i^2} z \\ &= k_{i+1} + (\sigma_{i+1}^2 - \sigma_i^2)s_\theta(k_{i+1},\sigma_{i+1}) + \sqrt{\sigma_{i+1}^2 - \sigma_i^2} z \end{aligned} \quad (5)$$

where $i$ is the total number of iterations, $z \sim N(0,1)$, $k(0) \sim p_0$, and we set $\sigma_0 = 0$ to simplify the notation.

In this work, we synergistically combine the generative network with conventional methods in our formulation, in which the iterative formulation can be transformed into:

$$\begin{aligned} k_i &= k_{i+1} + (\sigma_{i+1}^2 - \sigma_i^2)\nabla_k [\log p_t(k_{i+1}) + \log p_t(k_{con_{i+1}})] + \sqrt{\sigma_{i+1}^2 - \sigma_i^2} z \\ &= k_{i+1} + (\sigma_{i+1}^2 - \sigma_i^2)s_\theta(k_{i+1},\sigma_{i+1}) + (\sigma_{i+1}^2 - \sigma_i^2)\nabla_k \log p_t(k_{con_{i+1}}) \\ &\quad + \sqrt{\sigma_{i+1}^2 - \sigma_i^2} z \end{aligned}$$

(6)

where $\log p_t(k)$ is given by the prior model that represents information known beforehand about the true model parameter. $\log p_t(k_{con})$ is derived from low-rank data knowledge. The detailed introduction about the prior learning and reconstruction process of score-based model $s_\theta(k(t),t)$ is presented in Section III. C and Section III. D. Furthermore, the additional low-rank constraint $\log p_t(k_{con})$, which is determined by the traditional iterative algorithms (e.g., SAKE), is incorporated into the reconstruction architecture in a complementary and harmonious way. The detailed information is given in Section III. E.

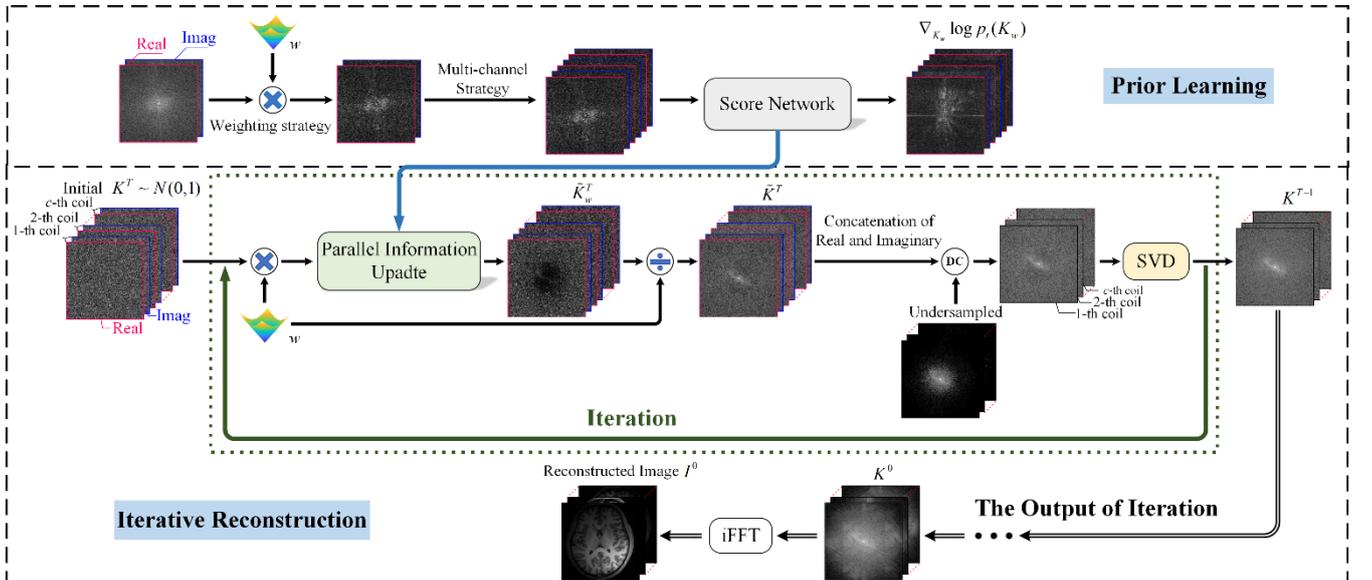

**Fig. 1.** Pipeline of the prior learning process and PI reconstruction procedure in WKGM. Top line: Prior learning is conducted in weight-k-space domain at a single coil image. Bottom line: PI reconstruction is conducted in iterative scheme that alternates between WKGM update and other traditional iterative

methods. It is worth noting that in our method, data consistency is performed after each predictor sampler and corrector sampler. For the brevity of the flowchart, we represent this process as performing data consistency after the PC sampler in the flowchart.

Fig. 1 intuitively shows the prior learning and reconstruction process of WKGM. In the prior learning stage, the complex-valued reference data are taken as the initial input of the training network. Subsequently, the weighting technology is enforced on each k-space input data, and the high-dimensional space augmentation design is imposed after the k-space weighting operation. Hence, we obtain a weighted high-dimensional tensor as the final input of the training network, where the real-value and imaginary-value components are concatenated along the channel dimension. Last, the score network is trained by learning the final composite k-space input data. For the reconstruction process, we use a PC sampler on the weighted input data to iteratively reconstruct the objects from the trained score network. The operation of removing the same weight is applied after PC sampling. In each iteration reconstruction step, the data consistency is simultaneously enforced to ensure that the output is consistent with the original k-space information. Moreover, the traditional k-space MRI algorithm (e.g., SAKE) is embedded to the reconstruction stage for better performance. Since WKGM is based on k-space domain, reconstruction in image domain is obtained by applying the inverse Fourier encoding. The final image is then obtained by the square root of the sum of squares (SOS).

### B. Motivation of Weight-k-Space Learning

In k-space, the low frequency and high frequency components are located in the central and surrounding regions, respectively. Since the k-space data exhibits a lot of variation in the dynamic range between low spatial frequency and high spatial frequency, training $s_\theta(k(t),t)$ directly on k-space domain is intractable. Inspired by the success of the works in [39]-[40], in this work, the weighting strategy is imposed on k-space input data to reduce the dynamic range between high frequency and low frequency information. The k-space weighting technology can be represented as:

$$k_w = wk, \quad w = (r \cdot x_k^2 + r \cdot y_k^2)^p \quad (7)$$

where $k$ denotes k-space input data of the training network, $w$ represents weight matrix. $k_w$ stands the weighted k-space input data. $x_k$ and $y_k$ are the count of frequency encoding lines and phase encoding lines, respectively. $r$ and $p$ stand two parameters for adjusting weight, where $r$ sets the cutoff value and $p$ determines the smoothness of the weight boundary.

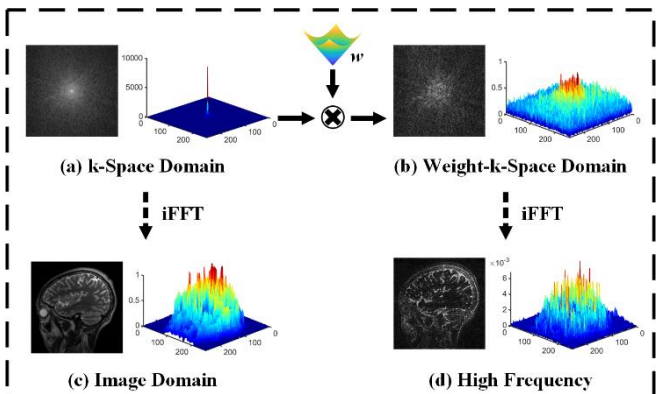

**Fig. 2.** Visual comparison of the amplitude images obtained from the k-space domain and weight-k-space domain, respectively. (a) The reference k-space data and its amplitude values. (b) The weight-k-space data and its amplitude values. (c) The image obtained by applying the inverse Fourier encoding on k-space data. (d) The image obtained by applying the inverse Fourier encoding on weight-k-space data.

As shown in Fig. 2, it can be seen that the amplitude values of pixel locations in weighted k-space domain are much closer and homogeneous. Meanwhile, it can be noticed that the main component of signals in the image support is low-frequency information, thus image support can also be efficiently reduced by suppressing low-frequency information with the weighting technique. Also, using weighting strategy, the corresponding output is high-frequency as feature, which is sparser and consistent with the spirit of compressed sensing to better MRI reconstruction. As shown in Fig. 3, the weighting technology is applied to multi-coil k-space data for MRI reconstruction in this work. Using k-space weighting techniques on multi-coil data reduces the dynamic range of each coil k-space data, so that the images show more relatively high frequency information.

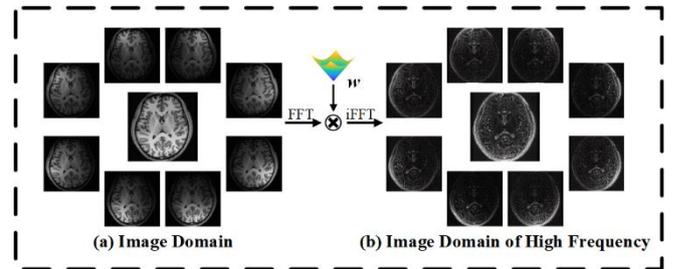

**Fig. 3.** The amplitude images before (a) and after (b) weighting operation.

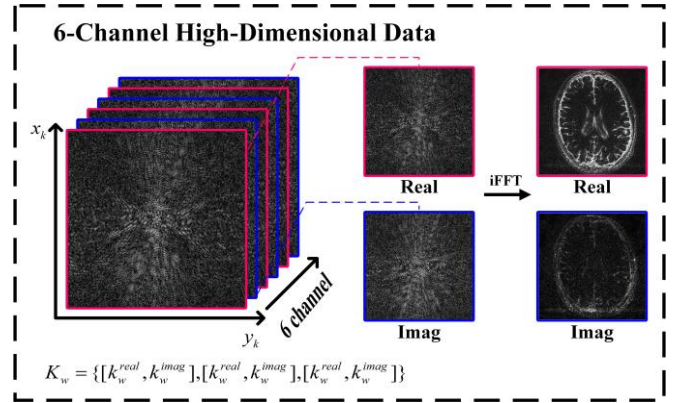

**Fig. 4.** Visualization of the 6-channel high-dimensional data $K_w$.

In image domain, Quan *et al*. [25] elaborated that prior information learned from high-dimensional tensor can achieve more effective MRI reconstruction than low-dimensional counterpart. In this study, we adopt this idea into k-space domain. As show in Fig. 4, to generate network $s_\theta(k(t),t)$ with prior data information at high-dimensional manifold and avoid difficulties for the accuracy in score estimation and sampling, we concatenate the real and imaginary components along the channel dimension, then form a 6-channel high-dimensional tensor $K_w = \{[k_w^{real}, k_w^{imag}], [k_w^{real}, k_w^{imag}], [k_w^{real}, k_w^{imag}]\}$ via variable augmentation technique. $[k_w^{real}, k_w^{imag}]$ denotes the channel concatenation of real and imaginary components. Eq. (6) can be represented as:

$$K_{w_i} = K_{w_{i+1}} + (\sigma_{i+1}^2 - \sigma_i^2)\nabla_{K_w}[\log p_t(K_{w_{i+1}}) + \log p_t(k_{con_{i+1}})] + \sqrt{\sigma_{i+1}^2 - \sigma_i^2}z$$

$$= K_{w_{i+1}} + (\sigma_{i+1}^2 - \sigma_i^2)s_\theta(K_{w_{i+1}},\sigma_{i+1}) + (\sigma_{i+1}^2 - \sigma_i^2)\nabla_{K_w}\log p_t(k_{con_{i+1}})$$

$$+ \sqrt{\sigma_{i+1}^2 - \sigma_i^2}z$$

s.t. $K_w = \{[k_w^{real}, k_w^{imag}],[k_w^{real}, k_w^{imag}],[k_w^{real}, k_w^{imag}]\}; \; k_w = wk$ (8)

where the weighted and high-dimensional k-space data $K_w$ is the final input data of the training network. Consequently, the conditional generation can be performed by alternatively iterative procedure among predictor-corrector (PC) sampling and low-rank penalty on Hankel matrix. In the following parts, we mainly introduce the solution process of these two subproblems.

### C. Prior Learning in Weight-k-Space

In score-based approaches, the diffusion process progressively adds noise to transform the complex data distribution into a Gaussian noise distribution from which we can sample as shown in Fig. 5. Song *et al.* formalized this noising process as the forward SDE shown in Eq. (3). In our work, it can be reformulated as:

$$dK_w = f(K_w,t)dt + g(K_w)dw \quad (9)$$

According to the work in Song *et al.*, we use the Variance Exploding (VE) SDE by choosing $f = 0, g = \sqrt{d[\sigma^2(t)/dt]}$ to form the following Markov chain:

$$K_{w_i} = K_{w_{i-1}} + \sqrt{\sigma_i^2 - \sigma_{i-1}^2}z_{i-1}, i = 1,\cdots,N \quad (10)$$

where $\sigma(t)$ is Gaussian noise function with a variable in continuous time $t \in [0,1]$, which can be redescribed as a positive noise scale $\{\sigma_i\}_{i=1}^N$. The process of Eq. (9) is reversed to transform Gaussian noise into data for samples generation. The reverse-time SDE can be given by:

$$dK_w = [f(K_w,t) - g(t)^2 \nabla_{K_w}\log p_t(K_w)]dt + g(t)d\overline{w} \quad (11)$$

The reverse-time SDE requires the time-dependent score function $\nabla_{K_w}\log p_t(K_w)$ to be known in advance. In practice, the score function $\nabla_{K_w}\log p_t(K_w)$ is intractable. Fortunately, it can be estimated by training a time-dependent score network $s_\theta(K_w(t),t)$ parameterized by $\theta$. $s_\theta(K_w(t),t)$ can be trained by optimizing the following objective:

$$\theta^* = \arg\min_\theta \mathbb{E}_t\{\lambda(t)\mathbb{E}_{K_w(0)}\mathbb{E}_{K_w(t)|K_w(0)}[$$
$$\|s_\theta(K_w(t),t) - \nabla_{K_w(t)}\log p_{0t}(K_w(t)|K_w(0))\|^2]\} \quad (12)$$

where $\lambda(t): \mathbb{R} \to \mathbb{R}$ is a weighting function. Vincent *et al.* [41] demonstrated that the minimizer of the objective $\theta^*$ will be such that $s_\theta(K_w(t),t) \approx \nabla_{K_w}\log p_t(K_w)$, which means that $\nabla_{K_w}\log p_t(K_w)$ is known for all $t$ by trained $s_\theta(K_w(t),t)$, allowing us to derive the reverse diffusion process of Eq. (9) and simulate it to sample.

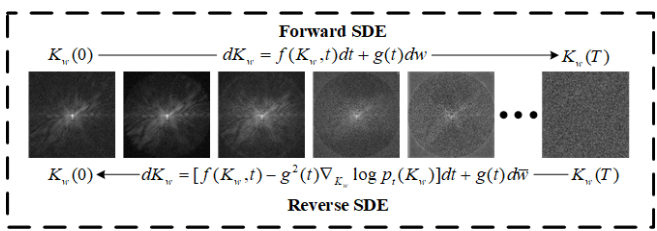

Fig. 5. Illustration of the forward and reverse processes of k-space data. The forward process can be accomplished with a continuous-time SDE. The reverse time SDE can be solved by training a score network to estimate the score function $\nabla_{K_w}\log p_t(K_w)$.

### D. PI Reconstruction of WKGM

Once score network $s_\theta(K_w(t),t)$ is trained, we insert it into Eq. (11) and solve the resulting reverse SDE to achieve MRI reconstruction:

$$dK_w = [f(K_w,t) - g(t)^2 s_\theta(K_w(t),t)]dt + g(t)d\overline{w} \quad (13)$$

To solve the reverse SDE, based on research by Song et al., the PC sampler is introduced in this work, where Variance Exploding (VE) SDE solver is selected as the predictor, annealed Langevin dynamics (ALD) as the corrector to correct the solution of a numerical VE SDE solver, as shown in Eq. (14). We can therefore obtain the samples from the obtained prior distribution by solving the reverse SDE numerically. Among them, each unconditional sampling update step involves the operation of weighting and dividing with the same weight in solving VE SDE and ALD sampling.

$$Predictor: K_{w_i} = K_{w_{i+1}} + (\sigma_{i+1}^2 - \sigma_i^2)s_\theta(K_{w_{i+1}},\sigma_{i+1}) + \sqrt{\sigma_{i+1}^2 - \sigma_i^2}z$$

$$Corrector: K_{w_i} = K_{w_{i+1}} + \varepsilon_i s_\theta(K_{w_{i+1}},\sigma_{i+1}) + \sqrt{2\varepsilon_i}z$$

(14)

where the initial solution $K_{w_t}$ can be a total uniform noise or other predefined value. $\varepsilon > 0$ is a step size, $i$ is the total number of iterations, and $z \sim N(0,1)$.

After the samples update step, weighted high-dimensional data $K_w$ is transformed to the high-dimensional tensor $K$ by removing the weight. And then $K$ is replaced by the k-space data $k_{gen}$ after channel averaging operation. Thus, we can obtain the subproblem regard to data consistency:

$$\min_k\{\|Mk - y\|_2^2 + \lambda \|k - k_{gen}\|_2^2\} \quad (15)$$

The data-consistency problem is solved via:

$$k(j) = \begin{cases} k_{gen}(j), & \text{if } j \notin \Omega \\ [k_{gen}(j) + \lambda y(j)]/(1+\lambda), & \text{if } j \in \Omega \end{cases} \quad (16)$$

where $\Omega$ denotes an index set of the acquired k-space samples. $k_{gen}(j)$ represents an entry at index $j$ in k-space. In the noiseless setting (i.e., $\lambda \to \infty$), we simply replace the $j_{th}$ predicted coefficient by the original coefficient if it has been sampled. After several iterations in k-space domain, the image is reconstructed by applying the inverse Fourier encoding $I = F^{-1}k(j)$. The final reconstruction is obtained by combining the channels through SOS. **Algorithm 1** compactly represents the training and sampling process of WKGM.

| Algorithm 1: WKGM |
|---|
| **Training stage** |
| 1: Dataset: Higher-dimensional weight-*k*-space dataset $\quad K_w = \{[k_w^{real}, k_w^{imag}],[k_w^{real}, k_w^{imag}],[k_w^{real}, k_w^{imag}]\}$ $\quad k_w = w \cdot k$ |
| 2: Output: Trained $S_\theta(K_w,\sigma)$ |
| **Reconstruction stage** |
| 1: $K_w^N \sim N(0,\sigma_T^2 I)$ |
| 2: For $i = N-1$ to $0$ do (Outer loop) |
| 3: $\quad \tilde{K}_w^i \leftarrow \text{Predictor}(K_w^{i+1},\sigma_i,\sigma_{i+1})$ |
| 4: $\quad \tilde{K}^i \leftarrow \tilde{K}_w^{i+1}/w$ |

| | |
|---|---|
| 5: | $\tilde{k}^i_{gen} = \text{Mean}(\tilde{K}^i)$ |
| 6: | Obtain $k^i$ by updating Eq. (16) |
| 7: | $\tilde{k}_w = w \cdot \tilde{k}^i$ |
| 8: | $K_w = \{[k_w^{real}, k_w^{imag}], [k_w^{real}, k_w^{imag}], [k_w^{real}, k_w^{imag}]\}$ |
| 9: | For $j=1$ to $M$ do (Inner loop) |
| 10: | $\tilde{K}_w^{i,j} \leftarrow \text{Corrector}(K_w^{i,j-1}, \sigma_i, \varepsilon_i)$ |
| 11: | $\tilde{K}_w^{i,j} \leftarrow \tilde{K}_w^{i,j}/w$ |
| 12: | $\tilde{k}_{gen}^{i,j} = \text{Mean}(\tilde{K}^{i,j})$ |
| 13: | Obtain $k^{i,j}$ by updating Eq. (16) |
| 14: | $\tilde{k}_w = w \cdot \tilde{k}^{i,j}$ |
| 15: | $K_w = \{[k_w^{real}, k_w^{imag}], [k_w^{real}, k_w^{imag}], [k_w^{real}, k_w^{imag}]\}$ |
| 16: | End for |
| 17: | End for |
| 18: | $k^{rec} \leftarrow \tilde{k}^0$ |
| 19: | Return $I_{sos} = \sqrt{\sum_{c=1}^{C}|(F^{-1}k_c^{rec})|^2}$ |

| | |
|---|---|
| 1: | $K_w^N \sim N(0, \sigma_T^2 I)$ |
| 2: | For $i = N-1$ to $0$ do (Outer loop) |
| 3: |  Step 3-8 in Algorithm 1 |
| 4: |  For $j=1$ to $M$ do (Inner loop) |
| 5: |   Step 10-14 in Algorithm 1 |
| 6: |   $A_k^j \leftarrow H(\tilde{k}^{i,j})$ (Hankel matrix) |
| 7: |   $[U, \Sigma, V] = svd(A_k^j)$ (Perform SVD) |
| 8: |   $G^{Th_r}(A_k) \leftarrow U(Th_r(diag(\Sigma)))V^\dagger$ |
| 9: |   $\tilde{k}_{con}^{i,j} \leftarrow h^\dagger(G^{Th_r}(A_k))$ |
| 10: |   Obtain $k^{i,j}$ by updating Eq. (16) |
| 11: |   Step 13-14 in Algorithm 1 |
| 12: |  End for |
| 13: | End for |
| 14: | Step 17-18 in Algorithm 1 |

### E. Combination with Traditional Algorithms

The present WKGM is designed in iterative manner. Hence, it offers flexibility for incorporating any traditional k-space domain algorithms into the whole reconstruction process. In this work, we combine WKGM with traditional algorithm SAKE [8] by alternating iteration reconstruction to further improve MRI reconstruction performance, and forms SVD-WKGM.

SAKE is a calibrationless reconstruction algorithm forming the whole k-space data into a Hankel structured low-rank matrix to recover images from undersampled multi-coil data. Specifically, SAKE connects the multi-coil data in series and the connected data is reformulated as the shape of the Hankel matrix used for Cadzow's signal enhancement. This process can be formalized as:

$$A_k = H(k) \quad (17)$$

where $A_k$ denotes the structured low-rank Hankel matrix. $H$ is the operator that generates data matrix $A_k$ from multi-coil dataset $k$ concatenated in a vector form and $H^\dagger$ is the corresponding pseudo-inverse operator, respectively. This structured matrix has low rank nature due to the linear dependency of the multi-coil MRI data [42]-[43]. As a result, the extension to recovery of PI without a fully sampled calibration region can be formulated as a structured low-rank matrix completion problem. The solution is obtained by a projection-onto-sets type algorithm with singular value thresholding, especially hard-thresholding [44].

$$G^{Th_r}(A_k) = U(Th_r(diag(\Sigma)))V^\dagger \quad (18)$$

where $G^{Th_r}$ represents singular values hard thresholding operator. $U\Sigma V^\dagger$ represents the SVD of $A_k$. Finally, we project the data matrix onto the k-space data. This operation is done implicitly by applying $H^\dagger$ to the data matrix. The process can be formulated as a constraint optimization, that is:

$$k_{con} = H^\dagger(G^{Th_r}(A_k)) \quad (19)$$

Overall, **Algorithm 2** exhibits the whole reconstruction procedure. It is worth noting that we only choose SAKE as an example to demonstrate the flexibility of WKGM. Other traditional algorithms like P-LORAKS [9] and ALOHA [10] can also be incorporated.

**Algorithm 2: Reconstruction stage in SVD-WKGM**

## IV. EXPERIMENTS

### A. Experimental setup

**1) Datasets:** The experiments are performed on six datasets: the *SIAT* dataset, the *T1-weighted* image, the *T1 GE Brain* dataset, the *T2 Transverse brain* dataset, the *T2-weighted brain* dataset and the *knee* dataset.

First, we use brain dataset for training. The brain dataset is selected from *SIAT* dataset, which is provided by Shenzhen Institutes of Advanced Technology, the Chinese Academy of Science. Informed consents are obtained from the imaging subject in compliance with the institutional review board policy. The collected dataset consists of 500 2D 12-channel complex-valued MR images with $256 \times 256$ pixels. These 12-channel complex-valued MR images are combined into single-coil images through coil compression (http://people.eecs.berkeley.edu/~mlustig/Software.html), and then augmented to 4000 single-coil images for training. In detail, MR images are acquired from a healthy volunteer on a 3.0T Siemens Trio Tim MRI scanner using the T2 weighted turbo spin echo sequence. Repetition time (TR)/echo time (TE) is $6100/99$ $ms$, and field of view (FOV) is $220 \times 220$ $mm^2$. Besides, the pixel size is $0.9 \times 0.9$ $mm^2$.

Aside from *SIAT* dataset, we conduct experiments on three in-vivo datasets with different sequences and contrasts to verify the generalization capability of WKGM. First, a brain image from a healthy volunteer is acquired with a T1-weighted, 3D spoiled gradient echo sequence. The FOV is $20 \times 20 \times 20$ $cm^3$. TR/TE is $17.6/8$ $ms$. Flip angle is $20°$. A single axial slice is selected from this data set and is used throughout the experiments. The scan is performed on a 1.5T MRI scanner (GE, Waukesha, Wisconsin, USA) using an 8-channel receive-only head coil. Second, the *T1 GE Brain* includes 8-channel complex-valued MR images with size of $256 \times 256$. The MR images are acquired by 3.0T GE. The FOV is $220 \times 220$ $mm^2$, and TR/TE is $700/11$ $ms$. Also, 12-channel *T2 Transverse brain* MR images with size of $256 \times 256$ are acquired with 3.0 T Siemens, whose FOV is $220 \times 220$ $mm^2$ and TR/TE is $5000/91$ $ms$.

To compare with the score-based image domain generative model, experiments are conducted on the 16-channel *T2-weighted brain* dataset from the NYU fastMRI [45], which is publicly available.

Finally, we use the k-space *knee* dataset

(http://mridata.org) to compare with the method of [16]. The raw data is acquired from 3D fast-spin-echo (FSE) sequence with proton density weighting included fat saturation comparison by a 3.0T whole body MR system (Discovery MR 750, DV22.0, GE Healthcare, Milwaukee, WI, USA). The number of coils is $8$. The TR/TE is $1550/25\ ms$. There are 320 slices in total, and the thickness of each slice is 0.6 $mm$. FOV is $160 \times 128\ mm^2$ and the size of acquisition matrix is $320 \times 256$.

**2) Parameter Settings:** In this subsection, parameter settings in the proposed algorithm are discussed. The batch size is set to 2 and Adam optimizer with $\beta_1 = 0.9$ and $\beta_2 = 0.999$ is utilized to optimize the network. For noise variance schedule, we fix $\sigma_{max} = 1$, $\sigma_{min} = 0.01$ and $r = 0.075$. For all the algorithms, we use $N = 1000$, $M = 1$ iterations. For the other parameters, we follow the settings in the work of Song et al. [23]. The training and testing experiments are performed with 2 NVIDIA TITAN GPUs, 12 GB.

**3) Evaluation Metrics:** To quantitatively evaluate the performance of the various reconstruction models, we choose two classic metrics, namely peak signal to noise ratio (PSNR), and structural similarity index measure (SSIM). The PSNR describes the relationship of the maximum possible power of a signal with the power of noise corruption, and the SSIM is used to measure the similarity between the original image and reconstructed images. For the convenience of reproducibility, the source code and representative results are available at: *https://github.com/yqx7150/WKGM.*

### B. Reconstruction Experiments

**1) Generalization performance.** To evaluate the generalization performance, we compare the proposed WKGM and SVD-WKGM methods with other state-of-the-art algorithms, including learning-free techniques ESPIRiT [6], SAKE [8], P-LORAKS [9], LINDBERG [46] and unsupervised method EBMRec [38]. We conduct the experiments under various sampling patterns (e.g., 2D Poisson, 2D Partial and 2D Random sampling) and different acceleration factors (i.e., 4×, 6×, 8×, and 10×).

Fig. 6 depicts the qualitative reconstruction results of WKGM and SVD-WKGM, ESPIRiT, P-LORAKS and SAKE, along with their corresponding 5× magnified residual images. We conduct the experiments on *T1-weighted* image under 2D Poisson sampling patterns with acceleration factors $R$=4, 8 and 2D Partial sampling patterns with acceleration factors $R$=3, 6, respectively. Notice that the test dataset contains 8-channel complex-valued MR images, while WKGM is trained on the *SIAT* dataset. From Fig. 6, we can observe that there are significant residual artifacts and amplified noise in the results obtained by ESPIRiT, SAKE, and P-LORAKS. Nevertheless, the reconstructed image of SVD-WKGM preserves more small textural details and has the least noise relative to the reference image. Furthermore, the difference images at the second row of Fig. 6 illustrate that SVD-WKGM achieves the lowest reconstruction errors. In ESPIRiT and P-LORAKS, there remains much more noise in their reconstruction results. In SAKE, the overall reconstruction errors are higher than those from SVD-WKGM. Table I presents the quantitative comparisons of these methods. The image reconstructed by SVD-WKGM has the largest PSNR and SSIM values than the other reconstructions under different sampling strategies and acceleration factors, indicating the effectiveness of the proposed method. These observations correspond well with the qualitative analysis shown in Fig. 6.

TABLE I
PSNR AND SSIM COMPARISON WITH STATE-OF-THE-ART METHODS UNDER DIFFERENT SAMPLING PATTERNS WITH VARYING ACCELERATE FACTORS.

| (a) | ESPIRiT | SAKE | P-LORAKS | WKGM | SVD-WKGM |
|---|---|---|---|---|---|
| 2D Poisson $R$=4 | 33.97/0.862 | 35.71/0.916 | 34.62/0.896 | 35.13/0.905 | **36.15/0.923** |
| 2D Poisson $R$=8 | 32.04/0.802 | 32.92/0.851 | 30.88/0.788 | 32.02/0.860 | **34.03/0.881** |
| (b) | ESPIRiT | SAKE | P-LORAKS | WKGM | SVD-WKGM |
| 2D Partial $R$=3 | 30.12/0.838 | 30.87/0.904 | 28.97/0.888 | 30.21/0.892 | **31.05/0.905** |
| 2D Partial $R$=6 | 29.68/0.810 | 30.18/0.862 | 28.37/0.826 | 28.74/0.829 | **30.35/0.865** |

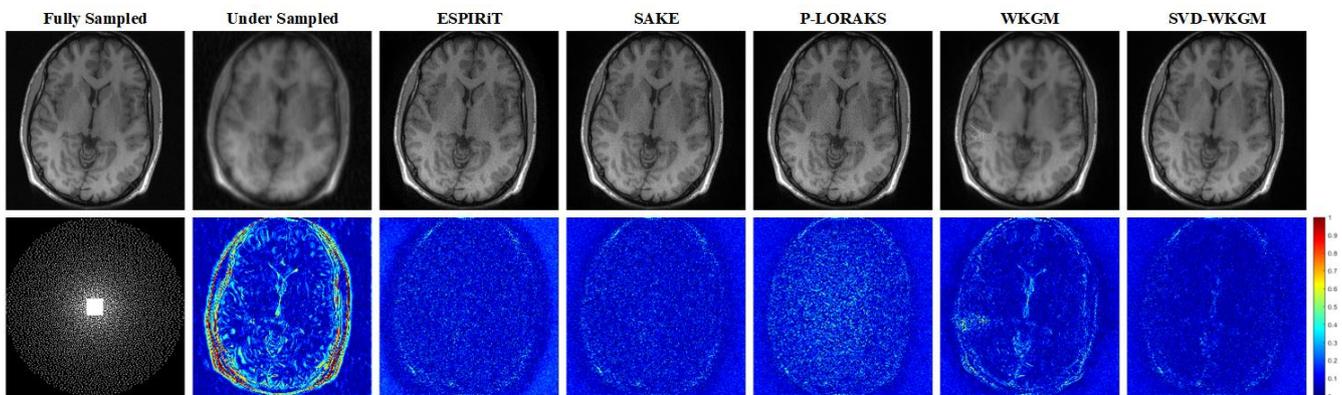

**Fig. 6.** Complex-valued PI reconstruction results at $R$=8 using 2D Poisson sampling with 8 coils. From left to right: Fully sampled, undersampled, reconstruction by ESPIRiT, SAKE, P-LORAKS, WKGM and SVD-WKGM. The intensity of residual maps is five times magnified.

Figs. 7 and 8 intuitively illustrate the representative reconstructions of the proposed methods and competitive methods, such as ESPIRiT, LINDBERG, EBMRec and SAKE. For *T1 GE Brain* dataset, we use 2D Random sampling patterns, with acceleration factors of 4 and 6, respectively. For *T2 Transverse Brain* dataset, the 2D Poisson sampling patterns with acceleration factors of 4 and 10 are applied to test. In line with previous results, SVD-WKGM provides better image quality with noise-like artifacts effectively suppressed compared with competitive results. The second row in Figs. 7 and 8 depicts the reconstruction errors of WKGM, SVD-WKGM, and compared methods. It indicates that ESPIRiT, EBMRec, and SAKE show noticeable noise-like residuals associated with the vulnerability of noise at high acceleration compared to the reference images. LINDBERG exhibits spatial blurring, which is effectively mitigated by SVD-WKGM reconstruction. It can also be seen that SVD-WKGM can generate superior performance, even at an extremely high acceleration factor, as shown in Fig. 8. Correspondingly, Table II tabulates the numerical results of the proposed methods and other comparison methods in terms of PSNR and SSIM metrics. In most cases, both PSNR and SSIM of WKGM are higher than ESPIRiT, LINDBERG and EBMRec, followed by SAKE. However, SVD-WKGM still outperforms the competitive SAKE.

It can also be observed that SVD-WKGM consistently outperforms other competitive methods under various sampling strategies. SVD-WKGM reduces sampling-specific effects, i.e., the difference between sampling schemes becomes less severe. These experiments demonstrate that the proposed model can be well generalized to various sampling patterns with different acceleration factors.

*2) Comparison with image domain score-based generative model.* To verify the superiority of WKGM, we perform comparison studies with csgm-mri-langevin [27] which is the image domain score-based generative model. Table III lists the quantitative comparisons of the proposed SVD-WKGM with zero-filled and csgm-mri-langevin [27], respectively. The experimental results are obtained on 16-channel *T2-weighted brain* image under Vertical and Horizontal sampling masks with acceleration factor *R*=3. Compared to csgm-mri-langevin, the reconstruction performance of SVD-WKGM improves with significantly higher PSNR and SSIM values under two different types of sampling patterns. Specifically, the PSNR value of the reconstructed image by SVD-WKGM increases up to 1 dB compared to csgm-mri-langevin. Fig. 9 depicts the qualitative results of SVD-WKGM and competition methods, it can be seen that SVD-WKGM has much lower errors with finer reconstruction details and much more edge information compared to csgm-mri-langevin. In general, SVD-WKGM produces visually more convincing and accurate reconstruction with higher PSNR and SSIM values.

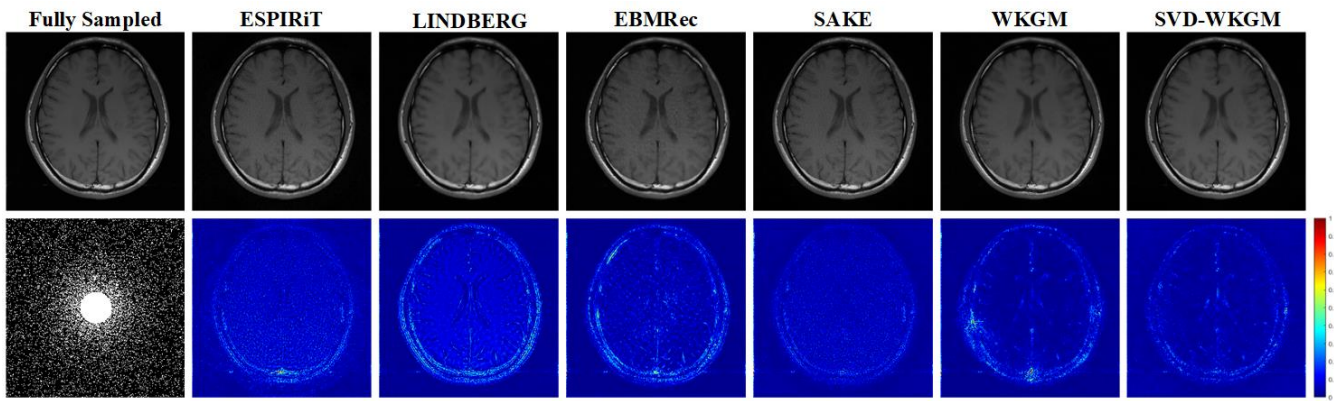

**Fig. 7.** PI results by ESPIRiT, LINDBERG, EBMRec, SAKE and SVD-WKGM on *T1 GE Brain* image at *R*=6 using a 2D Random sampling mask.

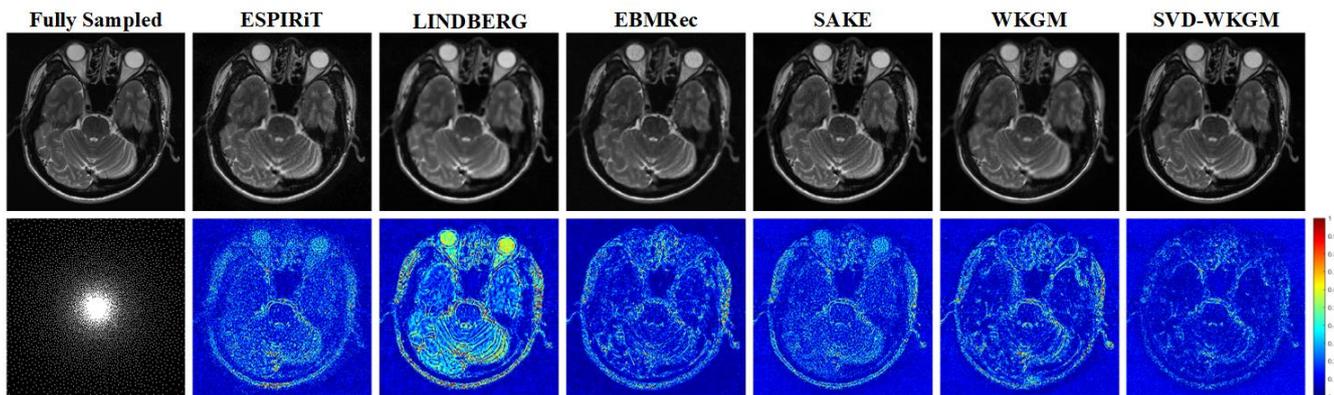

**Fig. 8.** PI reconstruction results by ESPIRiT, LINDBERG, EBMRec, SAKE, WKGM and SVD-WKGM on *T2 Transverse Brain* image at *R*=10 using 2D Poisson sampling mask. The intensity of residual maps is five times magnify.

TABLE II
PSNR AND SSIM COMPARISON WITH STATE-OF-THE-ART METHODS UNDER 2D RANDOM AND 2D POISSON SAMPLING PATTERNS WITH VARYING ACCELERATE FACTORS.

| *T1 GE Brain* | ESPIRiT | LINDBERG | EBMRec | SAKE | WKGM | SVD-WKGM |
|---|---|---|---|---|---|---|
| 2D Random *R*=4 | 39.08/0.933 | 38.98/0.961 | 40.17/0.968 | 41.54/0.952 | 40.67/0.969 | **43.85/0.970** |
| 2D Random *R*=6 | 36.01/0.921 | 35.16/0.958 | 36.55/0.952 | 38.09/0.932 | 37.14/0.957 | **39.94/0.960** |
| *T2 Transverse Brain* | ESPIRiT | LINDBERG | EBMRec | SAKE | WKGM | SVD-WKGM |
| 2D Poisson *R*=4 | 31.74/0.819 | 32.87/0.901 | 33.19/0.915 | 33.91/0.896 | 33.35/0.907 | **34.58/0.917** |
| 2D Poisson *R*=10 | 28.95/0.798 | 26.17/0.822 | 29.59/0.839 | 29.75/0.823 | 29.17/0.823 | **31.69/0.841** |

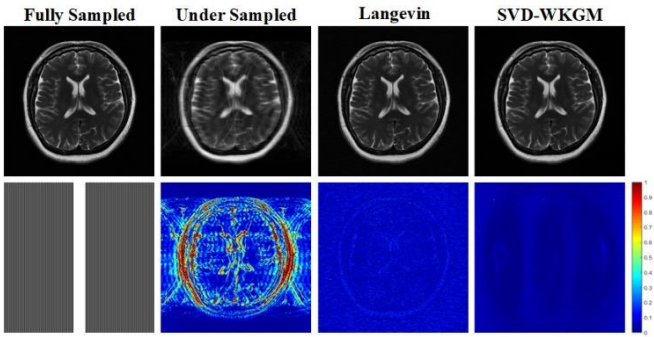

**Fig. 9.** PI reconstruction results by zero-filled, csgm-mri-langevin, and SVD-WKGM on *T2-weighted brain* image at *R*=3 vertical sampling mask. The intensity of residual maps is five times magnify.

TABLE III
PSNR AND SSIM COMPARISON WITH CSGM-MRI-LANGEVIN METHOD UNDER DIFFERENT PATTERNS AT R=3.

| T2 Brain | Zero-filled | Langevin | SVD-WKGM |
|---|---|---|---|
| Vertical *R*=3 | 23.51/0.693 | 37.47/0.940 | **38.42/0.953** |
| Horizontal *R*=3 | 26.28/0.759 | 38.44/0.950 | **39.51/0.960** |

**3)** *Comparison with the supervised k-space deep learning algorithm.* To further assess the reconstruction performance of SVD-WKGM, a comparison with the supervised k-space deep learning [16] is conducted on the *knee* dataset. The quantitative analysis of the reconstructed image in TABLE IV, shows PSNR and SSIM values of the reconstructed knee image using Cartesian trajectory at *R*=3 and 2D Poisson trajectory at *R*=6. As can be seen, the experimental result of the supervised k-space deep learning under *R*=3 Cartesian sampling trajectory has slightly higher PSNR and SSIM values than SVD-WKGM. It is reasonable because k-space deep learning is an end-to-end learning reconstruction method that requires training labels, while SVD-WKGM is completely unsupervised.

TABLE IV
PSNR AND SSIM COMPARISON WITH METHOD [16] UNDER DIFFERENT SAMPLING PATTERNS ON 8-COILS KNEE DATA.

| Cartesian *R*=3 | Zero-filled | [16] | SVD-WKGM |
|---|---|---|---|
| 100[th] Slice | 31.91/0.748 | **35.35/0.813** | 34.46/0.790 |
| 16[th] Slice | 31.52/0.749 | **35.99/0.836** | 33.88/0.805 |
| 2D Poisson *R*=6 | Zero-filled | [16] | SVD-WKGM |
| 100[th] Slice | 23.02/0.690 | 30.96/0.767 | **31.70/0.810** |
| 16[th] Slice | 20.48/0.643 | 28.99/0.793 | **29.81/0.832** |

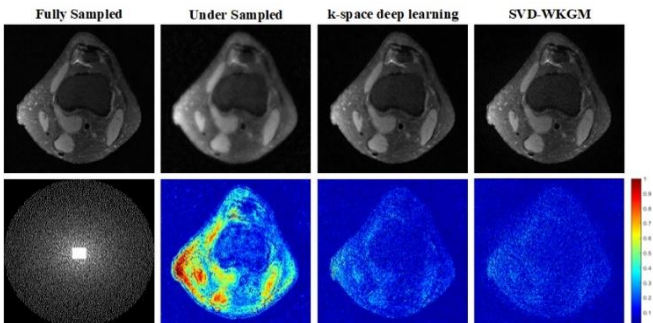

**Fig. 10.** PI reconstructions by zero-filled, k-space deep learning, and SVD-WKGM on 16[th] slice of 8 coils knee k-space data at *R*=6 using 2D Poisson sampling. The intensity of residual maps is four times magnify.

Nevertheless, under 2D Poisson sampling strategy with acceleration factor *R*=6, SVD-WKGM reconstructs image with higher PSNR and SSIM values compared to k-space deep learning method. Additionally, the qualitative reconstruction results further confirm the quantitative performance improvements. Based on the reconstructions and errors residuals shown in Fig. 10, it can be concluded that SVD-WKGM is superior to k-space deep learning in terms of reconstruction quality and structure details preservation. SVD-WKGM provides significant advantages for high acceleration factors, with largely improved reconstruction quality.

### C. Convergence Analysis

We experimentally investigate the convergence of No-Weight, WKGM and SVD-WKGM in terms of quantitative metric PSNR with respect to the number of iteration reconstruction. We randomly select an example of reconstructing the brain image using 2D Poisson sampling pattern with acceleration factor *R*=3. As shown in Fig. 11, both the model No-Weight of unweighting strategy and the model SVD-WKGM rise rapidly as iteration increases. Compared with No-Weight, the PSNR curves of WKGM and SVD-WKGM approximately converge at a higher PSNR level with relatively larger iteration number. Meanwhile, the PSNR curve of No-Weight fluctuates greatly in the early iteration.

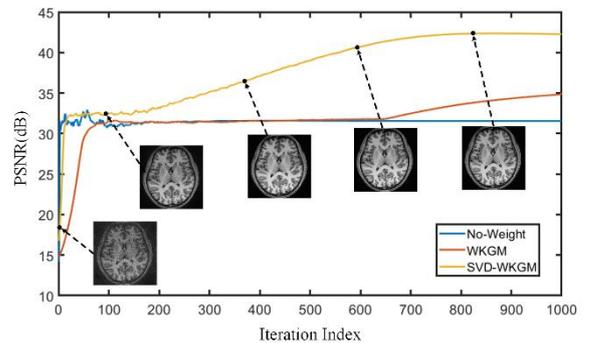

**Fig. 11**. Convergence curves of WKGM and SVD-WKGM in terms of PSNR versus the iteration number when reconstructing the brain image from 1/3 sampled data under 2D Poisson sampling pattern.

### D. Ablation Study

To demonstrate the model generalization capability and robustness, i.e., to verify our hypothesis that the proposed model can be trained on an arbitrary coil, which can be applied to the reconstruction of any coils without much performance loss, we perform an ablation study for PI reconstruction on 12 coils *SIAT* image using 2D Radial sampling pattern with acceleration factors *R*=3.3 and 5. Table V lists the quantitative comparisons of the models trained on different coils (e.g., 1-th, 11-th and single coil merged from 12 coils *SIAT* image). It can be found that the gap in reconstruction performance of the different coils trained models stays fairly small and constant.

TABLE V
PSNR AND SSIM COMPARISON OF DIFFERENT TRAINING MODEL UNDER DIFFERENT SAMPLING PATTERNS WITH VARYING ACCELERATE FACTORS ON 12-COILS SIAT DATA.

| WKGM | 1[th] Coil | 11[th] Coil | Single coil |
|---|---|---|---|
| 2D Radial *R*=3.3 | 34.33/0.920 | 34.34/0.922 | 34.32/0.921 |
| 2D Radial *R*=5 | 31.05/0.906 | 31.59/0.904 | 31.09/0.902 |

Table VI evaluates the impact of different values of weight parameters *r* and *p* on reconstructing *T1 GE Brain* image and the corresponding frequency range. The original range of *T1 GE Brain* image is 3820.8. One can observe that weight embedding strategy results in a significant reduction of dynamic range. We can therefore conclude that the appropriate selection of parameters *r* and *p* plays an important role in reconstruction.

To investigate the performance of WKGM under different channels, we visualize the reconstruction results of the models trained on different channel data in Fig. 12. All results are obtained with 1D Cartesian acquisition trajectories with acceleration factors *R*=3, 6.7 on *T1 GE Brain* image. From the reconstruction results and corresponding error maps, it can be seen that models trained on 6 and 8-channel data obtain better reconstruction results and lower errors than those of 2 and 4-channel counterparts. There is little difference between the reconstruction result of 6-channel and 8-channel models, which is also indicated by the variations of quantitative metrics in Table VII. Considering that model trained on 8-channel data needs much more computation than model trained on 6-channel data, the 6-dimensional space strategy is applied in all the experiments.

TABLE VI
PSNR AND SSIM COMPARISON OF DIFFERENT WEIGHT PARAMETER VALUES UNDER R=6 POISSION SAMPLING MASK AND CORRESPONDING FREQUENCY RANGE.

| WKGM | $p = 0.4$ | $p = 0.5$ | $p = 0.6$ |
|---|---|---|---|
| $r = 1 \times 10^{-2}$ | **38.47**/0.966 | 38.36/**0.971** | 32.24/0.874 |
| $r = 1.5 \times 10^{-2}$ | 37.50/0.940 | **38.39/0.973** | 34.13/0.936 |
| $r = 2 \times 10^{-2}$ | 35.65/0.935 | **38.54/0.972** | 36.20/0.963 |
| Frequency range after weighting | $p = 0.4$ | $p = 0.5$ | $p = 0.6$ |
| $r = 1 \times 10^{-2}$ | 2.24 | 0.39 | 0.07 |
| $r = 1.5 \times 10^{-2}$ | 3.09 | 0.58 | 0.11 |
| $r = 2 \times 10^{-2}$ | 3.89 | 0.78 | 0.15 |

TABLE VII
PSNR AND SSIM VALUES OF MODELS TRAINED ON DIFFERENT CHANNELS DATA UNDER 1D CARTESIAN SAMPLING.

| WKGM | 2ch | 6ch | 8ch |
|---|---|---|---|
| Cartesian $R$=3 | 34.87/0.953 | 35.73/**0.954** | **35.95/0.954** |
| Cartesian $R$=6.7 | 28.59/0.895 | 30.74/**0.904** | **30.79**/0.900 |

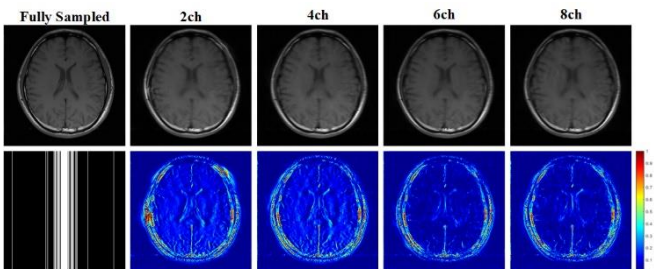

**Fig. 12.** PI reconstruction results of models trained on different channels data with *R*=6.7 Cartesian sampling pattern. From left to right: Zero-filled, 2-channel model, 4-channel model, 6-channel model, and 8-channel model.

Finally, we validate the effectiveness of weighting strategy. The model under No-Weight is trained without weighting strategy. Results of the assessment of No-Weight, WKGM and SVD-WKGM are shown in Fig. 13 and Table VIII. Unsurprisingly, better results can be achieved when applying WKGM and SVD-WKGM to test the *T2 Transverse Brain* data. Table VIII shows a large raise of PSNR and SSIM values under 2D Poisson sampling pattern with *R*=10. As can be seen in Fig. 13, WKGM and SVD-WKGM can fruitfully reconstruct the results with less error.

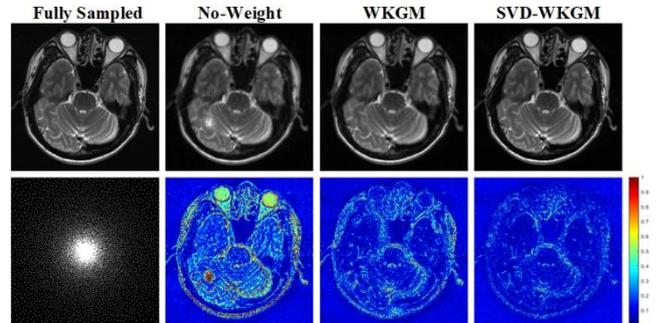

**Fig. 13.** PI reconstruction results by No-Weight, WKGM and SVD-WKGM on *T2 Transverse Brain* image at *R*=10 using 2D Poisson sampling mask.

TABLE VIII
PSNR AND SSIM COMPARISON WITH NO-WEIGHT METHOD UNDER 2D POISSON SAMPLING PATTERN.

| T2 Transverse Brain | No-Weight | WKGM | SVD-WKGM |
|---|---|---|---|
| 2D Poisson $R$=4 | 32.00/0.881 | 33.35/0.907 | **34.58/0.917** |
| 2D Poisson $R$=10 | 25.00/0.737 | 29.17/0.823 | **31.69/0.841** |

## V. CONCLUSION

This work proposed a weighted k-space generative model for multi-coil MRI reconstruction. The present WKGM applied a generative network and a weighting strategy to robust reconstruction, and incorporated traditional methods in a flexible way to improve PI reconstruction quality and implement calibration-less PI reconstruction. The calibration-less strategy minimized potential mismatches between calibration data and the main scan, while eliminating the need for fully sampled calibration region. Since k-space generative model could be used to directly interpolate missing k-space data, the distortion or disappearance of detailed structures caused by the zero-filled reconstruction could be avoid. The weighted k-space strategy made the model training to be more tractable and made better use of high frequency components. Experiment results verified that WKGM and SVD-WKGM can produce better reconstruction performance under different sampling patterns with large acceleration factors, keeping higher PSNR and SSIM values compared to state-of-the-arts. More traditional methods will be exploited in the future study to efficiently validate the algorithm flexibility.